\documentclass[12pt, twocolumn]{article}
\usepackage{graphicx}
\usepackage{amsmath}
\usepackage{float}
\usepackage{hyperref}
\usepackage{caption}
\usepackage{subcaption}
\usepackage{authblk}
\usepackage[capitalise]{cleveref}

\title{\textbf{Hierarchical Quantum Transport from Coupled Topological Domain-Wall States in SSH Chains}}

\author[1]{Alessio Palavicini}
\author[1]{César G. Galván}
\author[2]{Carlos Ramírez \thanks{corresponding author e-mail: carlos@ciencias.unam.mx}}
\affil[1]{Facultad de Ciencias, Universidad Autónoma de San Luis Potosí, Av. Chapultepec 3000, S. L. P., 78295, San Luis Potosí, México}
\affil[2]{Departamento de Física, Facultad de Ciencias, Universidad Nacional Autónoma de México, Apartado Postal 70542, Ciudad de México 04510, México}

\date{\today}

\begin{document}
\maketitle

\begin{abstract}
We investigate the transport properties of Su-Schrieffer-Heeger (SSH) chains containing multiple topological domain walls and show that their interaction generates a hierarchy of emergent spectral structures. Each domain wall contributes a localized state inside the SSH gap, and the hybridization of these states produces minibands whose signatures are directly reflected in the transmission spectra. By combining domain-wall lattices with different domain-wall separations, we construct effective SSH structures within the miniband subspace. The resulting transmission spectra reproduce the characteristic features of conventional SSH chains, including gap formation, finite-size resonances, and the correspondence between transmission spectra and band structure. The construction can be applied recursively, generating successive generations of effective SSH structures. As a consequence, effective SSH spectra repeatedly emerge within progressively narrower energy intervals, producing a self-similar hierarchy of minibands and spectral gaps. To understand the origin of this hierarchy, we develop an effective renormalized description based on successive decimation. The effective parameters exhibit a hierarchy of interlaced singularities whose number increases at each iteration. These singularities partition the energy axis into progressively finer intervals and provide a natural interpretation of the repeated fragmentation of the spectrum. Our results show that topological domain-wall states can act as emergent degrees of freedom from which multiscale transport channels, effective couplings, and hierarchical spectral structures may be engineered. More generally, the framework introduced here establishes a connection between recursive topological constructions, effective Hamiltonians, and the emergence of self-similar spectra in one-dimensional systems. 
\end{abstract}

\section{Introduction} \label{intro}

Topological materials have attracted considerable attention due to their ability to support robust states protected by topological invariants \cite{Su1979Solitons,KaneMele2005,Hasan2010,Qi2011TopologicalInsulators}. The existence of such states is intimately connected to the bulk-boundary correspondence, linking the topological properties of the bulk with the appearance of localized states at system boundaries or interfaces \cite{Asboth2016,Delplace2011,Qi2011TopologicalInsulators,Xie2023BulkBoundary}. Owing to their robustness against moderate perturbations and disorder \cite{Qi2011TopologicalInsulators,Oliveira2024Robustness}, topological states have been experimentally observed in a wide variety of physical platforms \cite{Konig2007,Yasuda2017,Lin2023TopologicalDefects} and are actively sought as potential foundations for technological devices \cite{Gruber2025TopologicalMaterials}.

Among the models used to investigate these phenomena, the Su-Schrieffer-Heeger (SSH) model has become a paradigmatic platform for studying topology in condensed matter. Originally introduced to describe polyacetylene \cite{Su1979Solitons,Su1980SSH,Heeger1988}, the SSH model captures the essential features of one-dimensional topological systems, including band-gap formation, topological phase transitions, and the emergence of localized edge states \cite{Su1980SSH,Asboth2016}. Because of its simplicity and versatility, it has served as a foundation for numerous studies addressing localization, transport, and topological state engineering \cite{Meier2016,Atala2013,mondal_su-schrieffer-heeger_2025}.

Beyond the well-known edge states of finite SSH chains, it has been shown that interfaces between topologically trivial and nontrivial regions can support localized domain-wall states \cite{Yasuda2017,munoz_topological_2018,Han2023}. An important feature of domain-wall states is that, unlike edge states, their position can be controlled through the design of the underlying lattice. This additional degree of freedom makes them attractive candidates for engineering transport properties and spectral structures through deliberately tailored topological defects. Their topological origin and spatial localization have motivated proposals ranging from protected transport channels to quantum-state transfer protocols \cite{Zurita2023FastTransfer,dangelis2020}. 

When neighboring domain walls are brought close enough for their localized states to overlap, the resulting hybridization splits the defect-state energies and gives rise to narrow bands inside the original SSH gap \cite{Perez2024DomainWallSSH,munoz_topological_2018,Song2026}. Consequently, additional transport channels become available within an energy region that would otherwise remain inaccessible in a conventional SSH chain \cite{Perez2024DomainWallSSH,Zurita2023FastTransfer}. The resulting transport properties are therefore determined not only by the bulk SSH bands but also by the collective behavior of the coupled domain-wall states.

The transport properties of SSH systems and related topological structures have been investigated through a variety of theoretical approaches \cite{Datta1995,mondal_su-schrieffer-heeger_2025,Diaz2025a,Diaz2025b}. In the present work, we employ the recursive scattering-matrix method \cite{li1996,rumpf2011,ramirez2017,ramirez2018}, which provides an efficient framework for obtaining the transport properties and band structure of tight-binding Hamiltonians and is particularly well suited for the study of large hierarchical structures. The recursive nature of the method closely parallels the recursive construction of the systems considered here, enabling successive generations of domain-wall structures to be incorporated through a modular procedure that remains computationally stable and efficient.

In this work, we investigate the transport properties of SSH chains containing multiple topological domain walls and show that their interaction generates a hierarchy of emergent spectral structures \cite{Song2026}. We first analyze the formation of domain-wall minibands and demonstrate how these states give rise to additional transport channels within the SSH gap. We then construct effective SSH structures within the miniband subspace and show that the characteristic spectral organization of the SSH model reappears at progressively smaller energy scales through a recursive construction. To provide a physical interpretation of this hierarchy, we develop an effective renormalized description based on successive decimation and show that the resulting effective parameters generate a hierarchy of interlaced singularities associated with the repeated formation of spectral gaps and minibands. 

The paper is organized as follows. In Sec.~\ref{method}, we introduce finite SSH chains and domain-wall lattices and analyze their transport properties. Section~\ref{Hier} presents the hierarchical SSH construction and the emergence of recursive miniband structures. In Sec.~\ref{renorm}, we develop the effective renormalized description and discuss the hierarchy of singularities generated through decimation. Finally, Secs.~\ref{disc} and \ref{conclusions} contain the discussion and conclusions.

\section{From SSH chains to domain-wall lattices} \label{method}
We begin by reviewing the transport properties of finite SSH chains and identifying the characteristic features of their transmission spectra. We then describe how domain walls can be generated within SSH lattices and examine the spectral and transport consequences of their interaction. The resulting domain-wall lattices constitute the building blocks of the recursive structures introduced in the following section.

\subsection{SSH chains}
\label{ssh}

\begin{figure}[tb]
    \centering
    \begin{subfigure}{\linewidth}
        \caption{}
        \label{fig:top_SSH}
        \vspace{-8pt} 
        \centering
        \includegraphics[width=\linewidth]{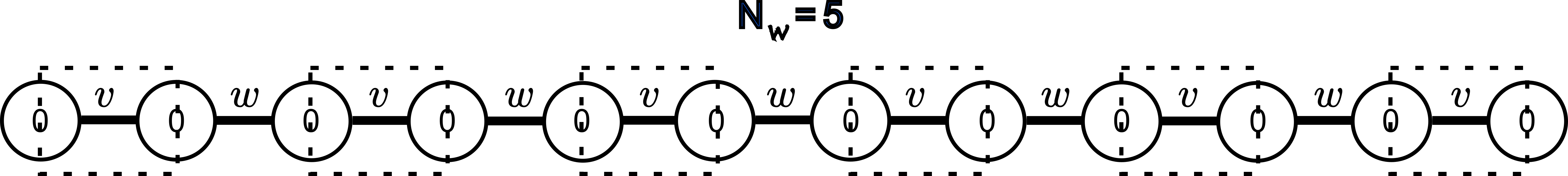}
        \vspace{-10pt}
    \end{subfigure}
    \begin{subfigure}{\linewidth}
        \caption{}
        \label{fig:bottom_SSH}
        \vspace{0pt} 
        \centering
        \includegraphics[width=\linewidth]{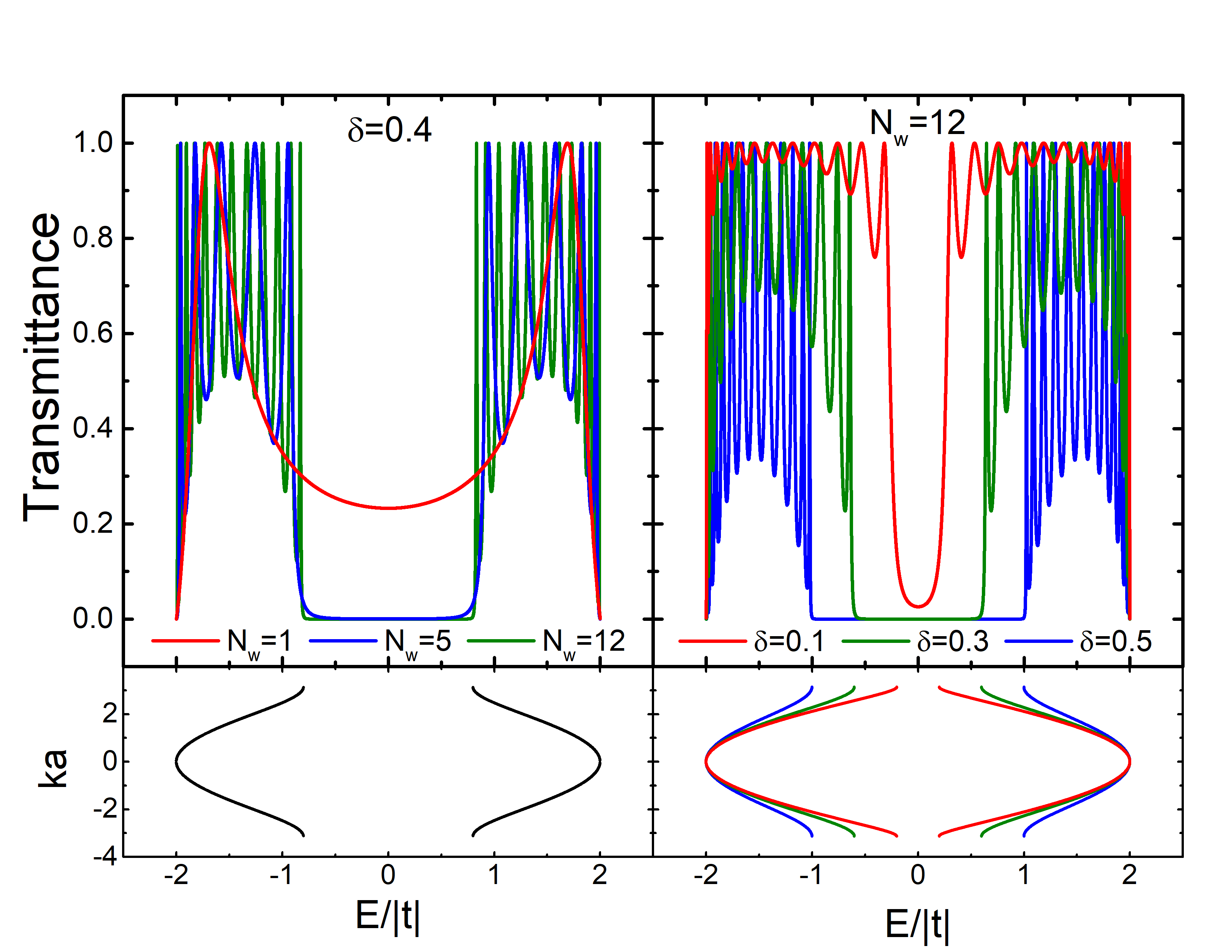}
    \end{subfigure}
    \vspace{-20pt}
    \caption{(a) Finite Su-Shrieffer-Heeger model with hopping amplitudes $v$ and $w$ and length $N_w=5$. (b) Transmittance and dispersion of SSH chains with: (Left) Fixed $\delta=0.4$ and $N_w=1$, $5$ and $12$, and (Right) Fixed length $N_w=12$, and $\delta=0.1$, $0.3$, and $0.5$. The number of transmission resonances increases with chain length, while the gap width is controlled by the dimerization parameter.}
    \label{fig1_SSH}
\end{figure}

The SSH model, depicted in \cref{fig:top_SSH}, consists of a one-dimensional lattice of sites with vanishing on-site energies. Its unit cell contains two sites connected through alternating hopping amplitudes $v$ (intracell) and $w$ (intercell). When $v\neq w$, the periodic system develops two energy bands separated by a gap. Throughout this work, the hopping amplitudes are parametrized as
\begin{equation}
v=1-\delta,
\qquad
w=1+\delta,
\end{equation}
where $\delta$ controls the degree of dimerization. For $\delta>0$ ($v<w$), the corresponding infinite SSH lattice lies in the non-trivial topological phase \cite{Su1980SSH}.

We consider finite SSH chains connected to uniform semi-infinite one-dimensional tight-binding leads having a hopping amplitude $t=1$, which sets the energy scale throughout this work. The finite SSH sequence always begins and ends with a $v$ hopping . The length of a chain is characterized by the number of $w$ bonds $N_w$. A chain with $N_w=5$ is shown in \cref{fig:top_SSH}.
\\
\Cref{fig:bottom_SSH} (left) shows the transmission spectra of SSH chains with fixed dimerization $\delta=0.4$ and lengths $N_w=1$, $5$, and $12$. \Cref{fig:bottom_SSH}(right) presents transmission spectra for a fixed length $N_w=12$ and dimerizations $\delta=0.1$, $0.3$, and $0.5$. The corresponding dispersion relations of the infinite periodic SSH lattices are shown below each spectrum.

Several characteristic features can be identified. First, the transmission spectra exhibit two transmission regions separated by a central gap, in agreement with the SSH band structure. However, for short chains, the transmission inside the gap does not vanish completely. This occurs because evanescent modes penetrating from the leads can still overlap across the finite scattering region, providing a finite transmission probability \cite{Datta1995}. For example, the case $N_w=1$ displays substantial transmission within the nominal gap, while for $N_w=12$, the overlap between evanescent states becomes negligible, and the transmission gap closely reproduces the forbidden region predicted by the bulk SSH dispersion relation.

A second important feature is the oscillatory structure within the transmission bands. The transmission maxima reach unity and their number is directly related to the chain length \cite{mondal_su-schrieffer-heeger_2025}. For instance, the chain with $N_w=5$ exhibits five dominant resonances within each transmission band. These resonances arise from the discrete finite-size states of the SSH chain.

The influence of the dimerization parameter is illustrated in \cref{fig:bottom_SSH}(right). Increasing $\delta$ widens the SSH gap and enhances the contrast between allowed and forbidden transmission regions. For $\delta=0.1$, the gap remains small, and evanescent coupling through the finite chain prevents the transmission from reaching zero. As $\delta$ increases, the transmission suppression becomes progressively stronger and approaches the behavior expected from the infinite periodic system. The case $\delta=0.5$ exhibits the correspondence between the transmission gap and the gap predicted by the SSH band structure.

The spectral fingerprint established in this subsection, namely a pair of transmission bands separated by a gap, together with a characteristic resonant structure inside each band, will provide the reference pattern for the hierarchical constructions discussed below.

\subsection{Domain-wall lattices}
\label{dws}

\begin{figure}[!tb]
    \centering
    \begin{subfigure}{\linewidth}
        \caption{}
        \vspace{-6pt}
        \centering
        \includegraphics[width=1.0\linewidth]{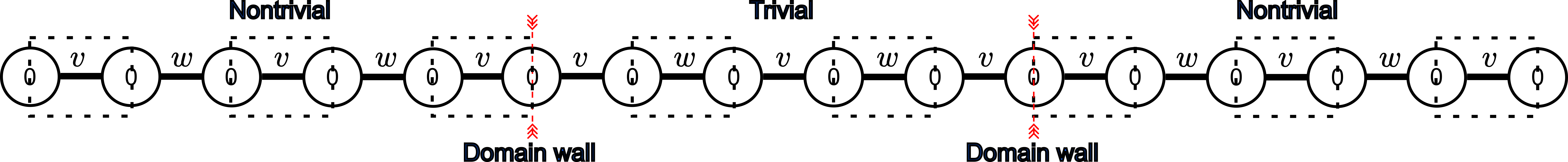}
        \vspace{-15pt}
        \label{fig_PDW}
    \end{subfigure}
    
    \begin{subfigure}{\linewidth}
        \caption{}
        \vspace{-2pt}
        \centering
        \includegraphics[width=1.0\linewidth]{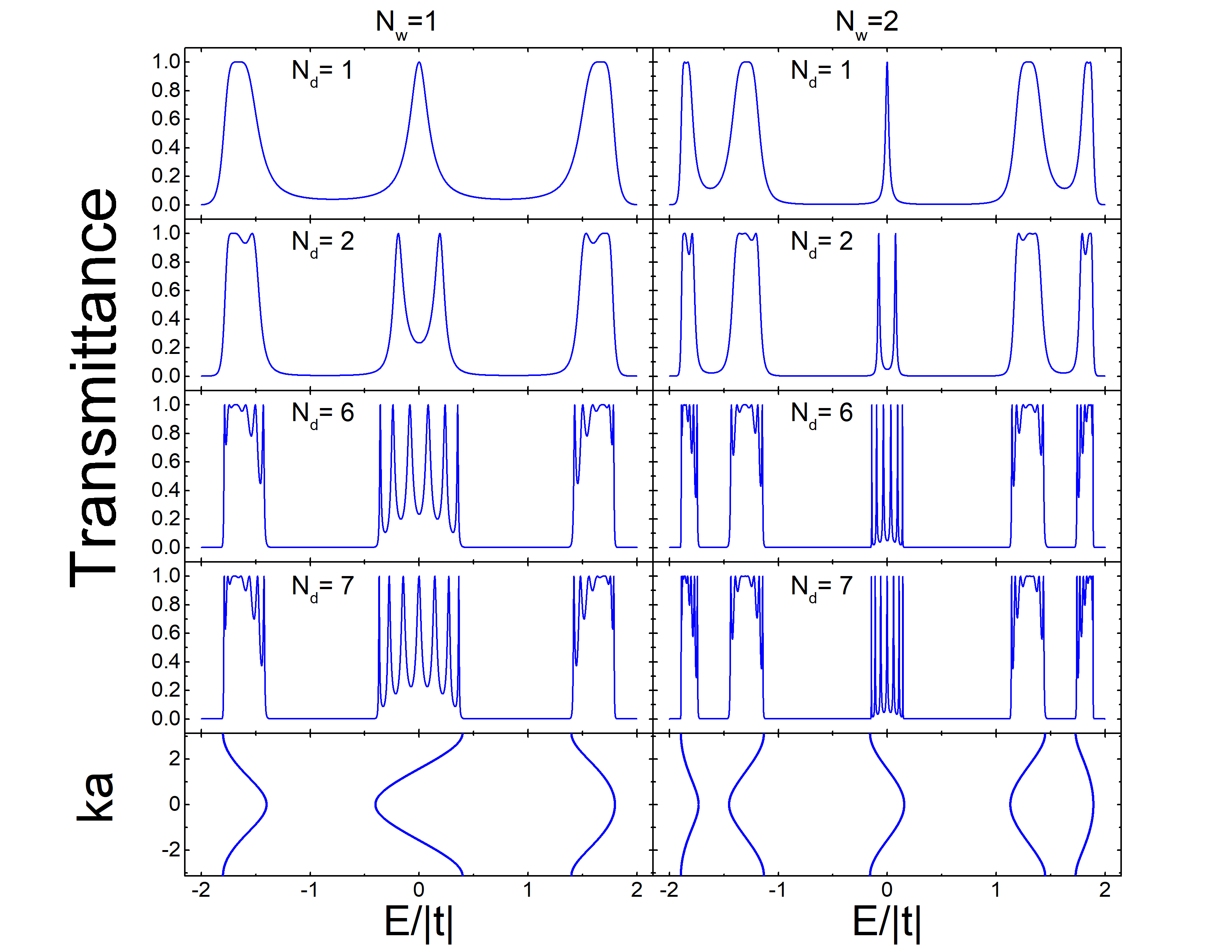}
        \label{fig2_res_DW}
    \end{subfigure}
    \vspace{-30pt}
    \caption{ Formation of minibands from coupled domain-wall states. (a) Domain-wall lattice obtained by concatenating SSH chains. The red dashed lines indicate the domain-wall positions. (b) Transmission spectra for increasing numbers of domain walls $N_d$ and the corresponding band structures of the associated infinite periodic systems. Hybridization of the localized domain-wall states produces minibands embedded within the original SSH gap. The cases $N_w=1$ and $N_w=2$ generate three-band and five-band spectra, respectively. }
    \label{fig3_PDW}
\end{figure}

Domain walls are introduced by joining finite SSH chains that begin and end with a hopping amplitude $v$. \Cref{fig_PDW} shows a structure obtained by concatenating finite SSH chains with $N_w=2$. The red dashed lines indicate the locations of the domain walls generated by the repeated $v-v$ hoppings \cite{munoz_topological_2018}. The dotted boxes highlight the intracell hopping amplitudes used to define the local unit-cell convention. The central region is characterized by a shifted unit-cell convention, leading to the opposite dimerization pattern and, therefore, a different topological character. The example shown in \cref{fig_PDW} consists of three concatenated SSH chains and thus contains two domain walls ($N_d=2$).

The transport signatures associated with these structures are presented in \cref{fig2_res_DW}. The left column corresponds to lattices built from SSH chains with $N_w=1$, while the right column corresponds to $N_w=2$, the configuration shown at the top of \cref{fig_PDW}. The successive rows show increasing numbers of domain walls. The bottom row presents the band structure of the corresponding infinite periodic systems.

Before discussing the domain-wall lattices, it is instructive to recall the behavior of an isolated finite SSH chain. As shown in \cref{ssh}, finite SSH chains exhibit a transmission gap around $E=0$. Although topological edge states exist at this energy when $v<w$, their wavefunctions decay exponentially into the bulk, and their overlap becomes negligible for sufficiently long chains. Consequently, no resonant transmission peaks appear inside the SSH gap when domain walls are absent.

The situation changes dramatically when a domain wall is introduced. For $N_d=1$, a single transmission resonance appears inside the gap. This resonance originates from the localized state associated with the domain wall and provides a transport channel across an energy region that would otherwise remain forbidden. Increasing the number of domain walls introduces additional localized states within the gap. The number of resonances observed in the transmission spectra coincides exactly with the number of domain walls ($N_d$), indicating that each domain wall contributes one localized state to the defect subspace. As their number increases, these localized states begin to hybridize and the corresponding resonances spread over a finite energy interval, eventually forming a miniband embedded within the original SSH gap \cite{munoz_topological_2018,Zurita2023FastTransfer}. This behavior is particularly evident for the cases with $N_d=6$ and $N_d=7$, where the resonances densely populate the same energy region occupied by the additional band observed in the corresponding periodic band structures. The excellent agreement between the miniband boundaries and the band structure calculations confirms that these new transport channels originate from the collective interaction of the domain-wall states.

The influence of the unit cell size can be observed by comparing the $N_w=1$ and $N_w=2$ cases. For $N_w=1$, the periodic structure possesses a three-site unit cell, leading to three allowed bands: the two SSH bulk bands and an additional miniband located inside the original gap. In contrast, the $N_w=2$ structures possess a five-site unit cell. The narrower central miniband observed for $N_w=2$ reflects the larger separation between neighboring domain walls. Besides the central miniband, the original SSH transmission bands split into additional subbands, resulting in a total of five allowed bands. This behavior is consistent with the enlarged unit cell and demonstrates that periodic arrays of domain walls modify not only the gap region but also the overall spectral organization of the system.

The results of this subsection emphasize that domain-wall lattices generate well-defined minibands whose spectral signatures are directly reflected in the transport properties. These minibands constitute the first level of spectral organization beyond the conventional SSH chain and provide the starting point for the constructions introduced in the next section.

\section{Hierarchical SSH construction} \label{Hier}

In this section we show how the domain-wall lattices introduced previously can be recursively combined to generate higher-order spectral structures.

\subsection{SSH minibands} \label{miniSSH} 

The domain wall lattices introduced in \cref{dws} can be used as templates to obtain hierarchical spectra. The central idea is to combine domain wall lattices with different separations between neighboring domain walls, producing distinct effective couplings between the corresponding localized states. These effective couplings can then be arranged to follow any pattern. 

\begin{figure}[!tb]
    \centering
    \begin{subfigure}{\linewidth}
        \caption{}
        \centering
        \includegraphics[width=\linewidth]{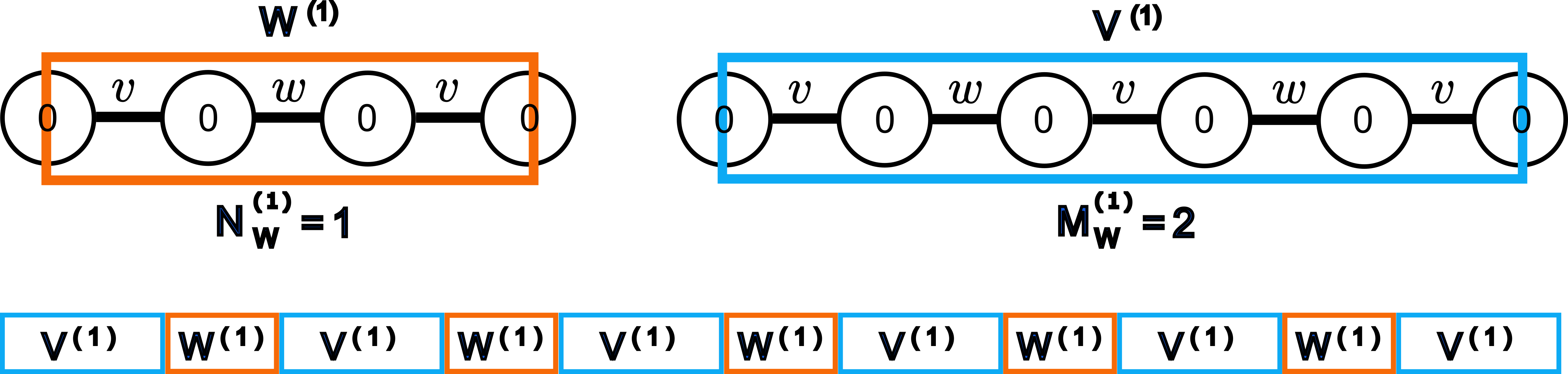}
        \vspace{-10pt}
        \label{fig_DW cells}
    \end{subfigure}
    
    \begin{subfigure}{\linewidth}
        \caption{}
        \centering
        \includegraphics[width=\linewidth]{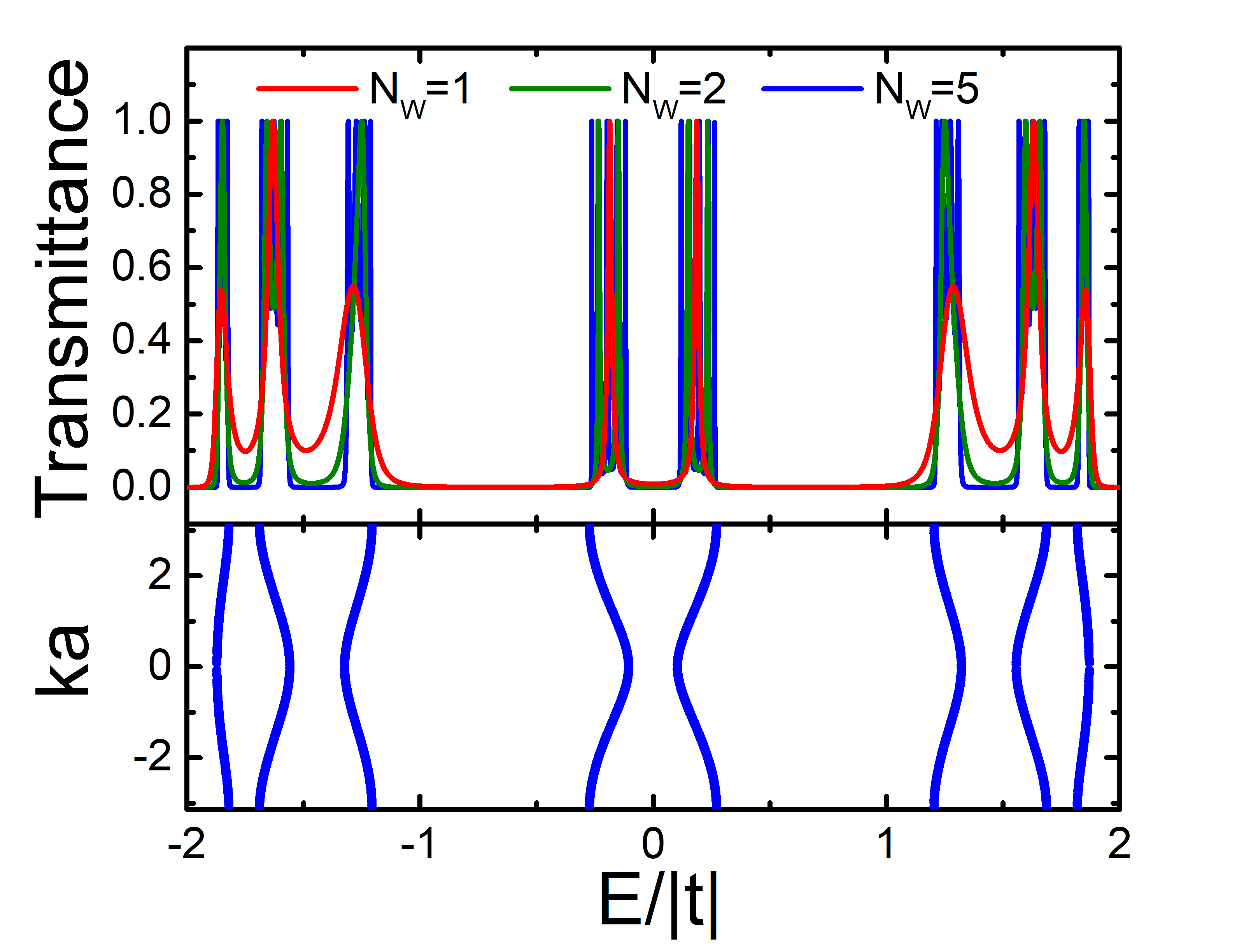}
        \label{fig3_res_SSHF}
        \vspace{-25pt}
    \end{subfigure}
    \label{fig_SSHF1}
    \caption{ (a) First-generation building blocks $W^{(1)}$ and $V^{(1)}$, corresponding to $N_W^{(1)}=1$ and $M_W^{(1)}=2$, respectively. The lower schematic illustrates the construction of an effective SSH chain from alternating $V^{(1)}$ and $W^{(1)}$ blocks. (b) Transmission spectra for effective SSH chains with $N_W=1$, $2$, and $5$, together with the band structure of the corresponding infinite periodic sequence. Calculations were performed for $\delta=0.4$. }
    
\end{figure}

Here we adopt the same alternating sequence that defines the SSH model. Although many choices are possible, throughout this work we focus on the simplest realization of this construction. We define two first-generation building blocks, named $W^{(1)}$ and $V^{(1)}$, consisting of finite SSH chains of different sizes. The number of SSH $w$ bonds contained in $W^{(1)}$ and $V^{(1)}$ is denoted by $N_W^{(1)}$ and $M_W^{(1)}$, respectively. In the examples considered in this work,

\begin{equation} 
    N_W^{(1)}=1 
    \qquad
    \text{and}
    \qquad
    M_W^{(1)}=2. 
\end{equation} 
As shown in \cref{fig_DW cells}, these structures correspond to 

\begin{equation} 
    W^{(1)}=vwv 
    \quad
    \text{and}
    \quad
    V^{(1)}=vwvwv. 
\end{equation}

The separation between neighboring domain walls is smaller in $W^{(1)}$ than in $V^{(1)}$. Consequently, the overlap between the corresponding domain-wall states is stronger in $W^{(1)}$, leading to a larger effective coupling between such states. In this sense, the structures $W^{(1)}$ and $V^{(1)}$ play analogous roles to two distinct effective couplings connecting neighboring domain-wall states. 

Using these building blocks, we construct finite chains following the same alternating pattern employed in the SSH model. The smallest first-generation structure is therefore 
\begin{equation} 
    V^{(1)}W^{(1)}V^{(1)}.
\end{equation}
Longer chains are obtained by repeating the same pattern and always starting and ending with a $V^{(1)}$ block. The size of the resulting structure is characterized by the number of $W^{(1)}$ blocks, denoted by $N_W$. 

The transmission spectra obtained from these structures are shown in \cref{fig3_res_SSHF} for $N_W=1$, $2$, and $5$. The lower panel displays the corresponding band structure of the infinite periodic sequence 

\[ \ldots V^{(1)}W^{(1)}V^{(1)}W^{(1)}V^{(1)}\ldots . \]

The periodic unit cell contains eight sites, corresponding to concatenating one $V^{(1)}$ block and one $W^{(1)}$ block. Therefore, the resulting dispersion relation consists of eight allowed bands, in agreement with the transport calculations. 

Several remarkable similarities with the conventional SSH chain can be identified. Most notably, the miniband occupying the central gap of the structures discussed in \cref{dws} now develops an internal gap and splits into two well-defined subbands. This behavior closely resembles the formation of the two SSH transmission bands described in \cref{ssh}, but now occurs within a considerably smaller energy interval. In other words, the characteristic spectral organization of the SSH model reappears inside the miniband generated by the domain-wall lattice. 

The finite-size effects displayed by these higher-order structures also follow the same trends observed in conventional SSH chains. For $N_W=1$, the transmission remains finite inside the newly generated gaps due to the overlap of evanescent modes across the finite structure. The same effect is still visible, although significantly reduced, for $N_W=2$. For $N_W=5$, the transmission spectrum closely reproduces the allowed and forbidden energy regions predicted by the infinite periodic band structure. 

A second hallmark of the effective SSH structure is found in the resonant structure of the transmission bands. Each of the newly generated minibands contains $N_W$ dominant transmission resonances, directly analogous to the $N_w$ resonances observed in finite SSH chains. The number of resonances therefore scales with the length of the effective structure, in direct analogy with the behavior observed for conventional SSH chains. Thus, the same finite-size spectral fingerprint identified in \cref{ssh} reappears at a smaller energy scale within the miniband spectrum. 

\subsection{Recursive construction}
\label{recursive} 

\begin{figure*}[!ht]
    \centering    
        \includegraphics[width=1\textwidth]{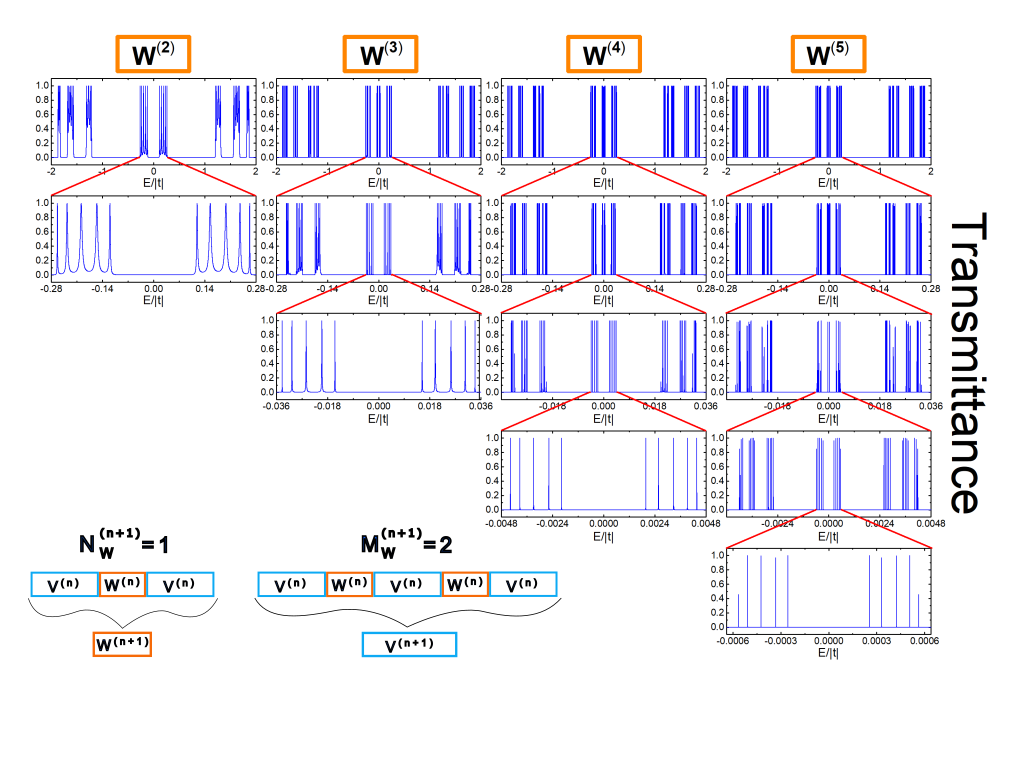}
        \vspace{-60pt}
        \caption{Hierarchical SSH structures and their transmission spectra. Hierarchy levels are shown from left to right and their corresponding progressive amplifications from top to bottom. Boxes: Effective SSH building blocks of an arbitrary hierarchical level $n$.}
        \label{fig4_res_Gens}
\end{figure*}

The SSH minibands discussed above constitute only the first step of a hierarchical construction. Once the first-generation building blocks $W^{(1)}$ and $V^{(1)}$ have been defined, the same procedure can be applied recursively to generate successive generations of structures.

The recursive rule employed throughout this work is illustrated in Fig.~\ref{fig4_res_Gens}. At hierarchy level $n$, two new building blocks, $W^{(n+1)}$ and $V^{(n+1)}$, are constructed from sequences of the previous-generation blocks $W^{(n)}$ and $V^{(n)}$. The sizes of these structures are determined by the parameters $N_W^{(n+1)}$ and $M_W^{(n+1)}$, which specify the number of $W^{(n)}$ blocks contained in $W^{(n+1)}$ and $V^{(n+1)}$, respectively.

The lower part of Fig.~\ref{fig4_res_Gens} illustrates the simplest realization of this recursive rule, corresponding to \begin{equation} N_W^{(n+1)}=1, \qquad M_W^{(n+1)}=2. \end{equation} In this case, \begin{equation} W^{(n+1)} = V^{(n)}W^{(n)}V^{(n)}, \end{equation} while $V^{(n+1)}$ is constructed from a longer sequence containing two $W^{(n)}$ blocks.

Repeated application of this rule produces a hierarchy of structures characterized by progressively smaller energy scales. Examples of the building blocks $W^{(N)}$ for hierarchy levels $N=2$, $3$, $4$, and $5$ are shown at the top of Fig.~\ref{fig4_res_Gens}. The corresponding transmission spectra are displayed below each structure. For all the cases, the recursive construction was performed using $N_W^{(k)}=1$ and $M_W^{(k)}=2$ for $k<N$, while the highest level was chosen with $N_W^{(N)}=5$.

The first row presents the transmission spectra over the full energy interval $-2\leq E\leq2$ for $\delta=0.4$. The remaining rows show successive magnifications of the central region of the preceding spectrum. As the hierarchy level increases, new transmission gaps emerge inside previously existing minibands. At the same time, each magnification reveals a spectral structure closely resembling the transmission spectrum observed at the preceding level.

At every hierarchical level, effective SSH transmission spectra reappear within progressively narrower energy intervals. In particular, the deepest magnifications recover the same qualitative spectral structure discussed in Sec.~\ref{ssh}, namely two transmission regions separated by a gap and containing a finite number of resonant peaks. For the examples shown here, each miniband contains five dominant resonances, reflecting the choice $N_W^{(N)}=5$. These results demonstrate that the hierarchical construction preserves the spectral organization of the parent structure at every generation, yielding a self-similar spectral hierarchy in which effective SSH spectra re-emerge on progressively reduced energy scales.

\section{Effective renormalized description}
\label{renorm}

The hierarchical spectra discussed in the previous sections originate from the interaction of localized domain-wall states. In Sec.~\ref{miniSSH}, we exploited the fact that domain walls separated by different distances generate different effective couplings, allowing the construction of effective SSH structures within the miniband subspace. To understand the microscopic origin of these couplings and the resulting spectral hierarchy, we develop an effective renormalized description based on the successive decimation of finite SSH segments.

\subsection{Effective description of domain-wall lattices}

\begin{figure*}[t] 
    \centering 
    \begin{minipage}[t]{0.3\textwidth}
        \centering 
        
        \begin{subfigure}{\linewidth} 
            \centering
            \vspace{-20pt}
            \caption{}
            \includegraphics[width=\linewidth]{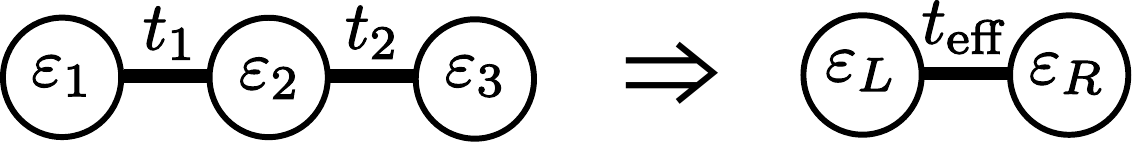}  
            \label{fig_renorm_top} 
        \end{subfigure} 
        
        \begin{subfigure}{\linewidth}
            \centering
            \vspace{5pt} 
            \caption{}
            \includegraphics[width=\linewidth]{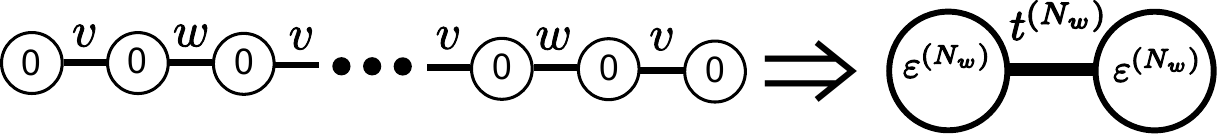}
            \label{fig_renorm_bottom} 
        \end{subfigure}

        \begin{subfigure}{\linewidth}
            \centering
            \vspace{5pt} 
            \caption{}
            \includegraphics[width=\linewidth]{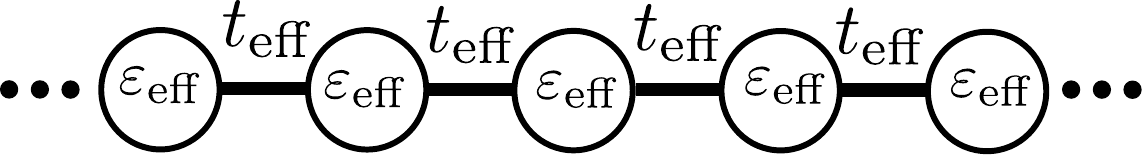}
            \label{fig_periodica_renorm} 
        \end{subfigure}
    \end{minipage} 
    \begin{minipage}[t]{0.33\textwidth}
        \vspace{-50pt}
        \centering
        \begin{subfigure}{\linewidth}
            \centering
            \caption{}
            \includegraphics[ width=\linewidth, trim={3cm 0.5cm 3.6cm 1.8cm}, clip ]{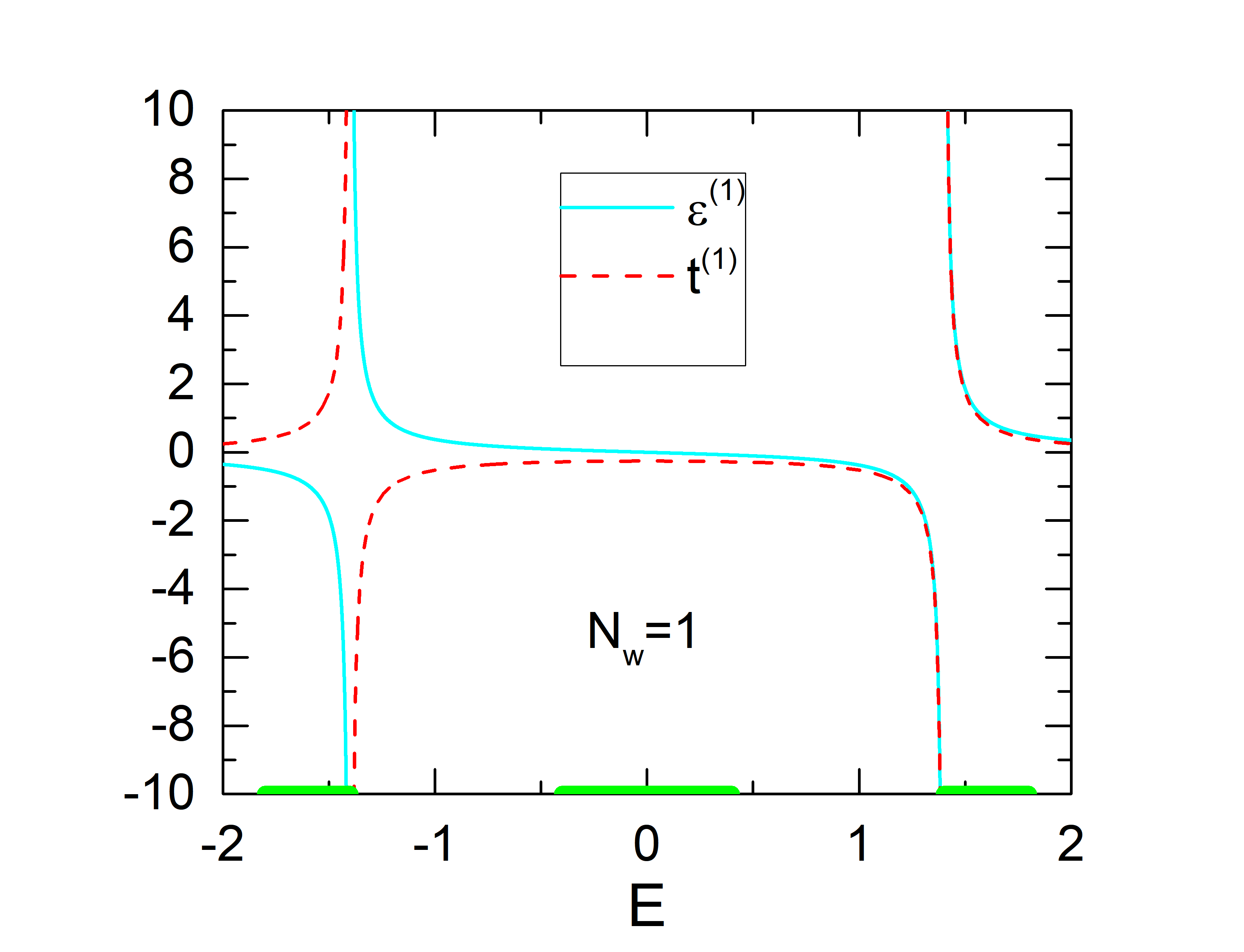}
            \label{fig_eff_Nw1}
        \end{subfigure} 
    \end{minipage} 
    \hfill 
    \begin{minipage}[t]{0.33\textwidth}
        \vspace{-50pt}
        \centering
        \begin{subfigure}{\linewidth}
            \centering
            \caption{}
             \includegraphics[ width=\linewidth, trim={3cm 0.5cm 3.6cm 1.8cm}, clip ]{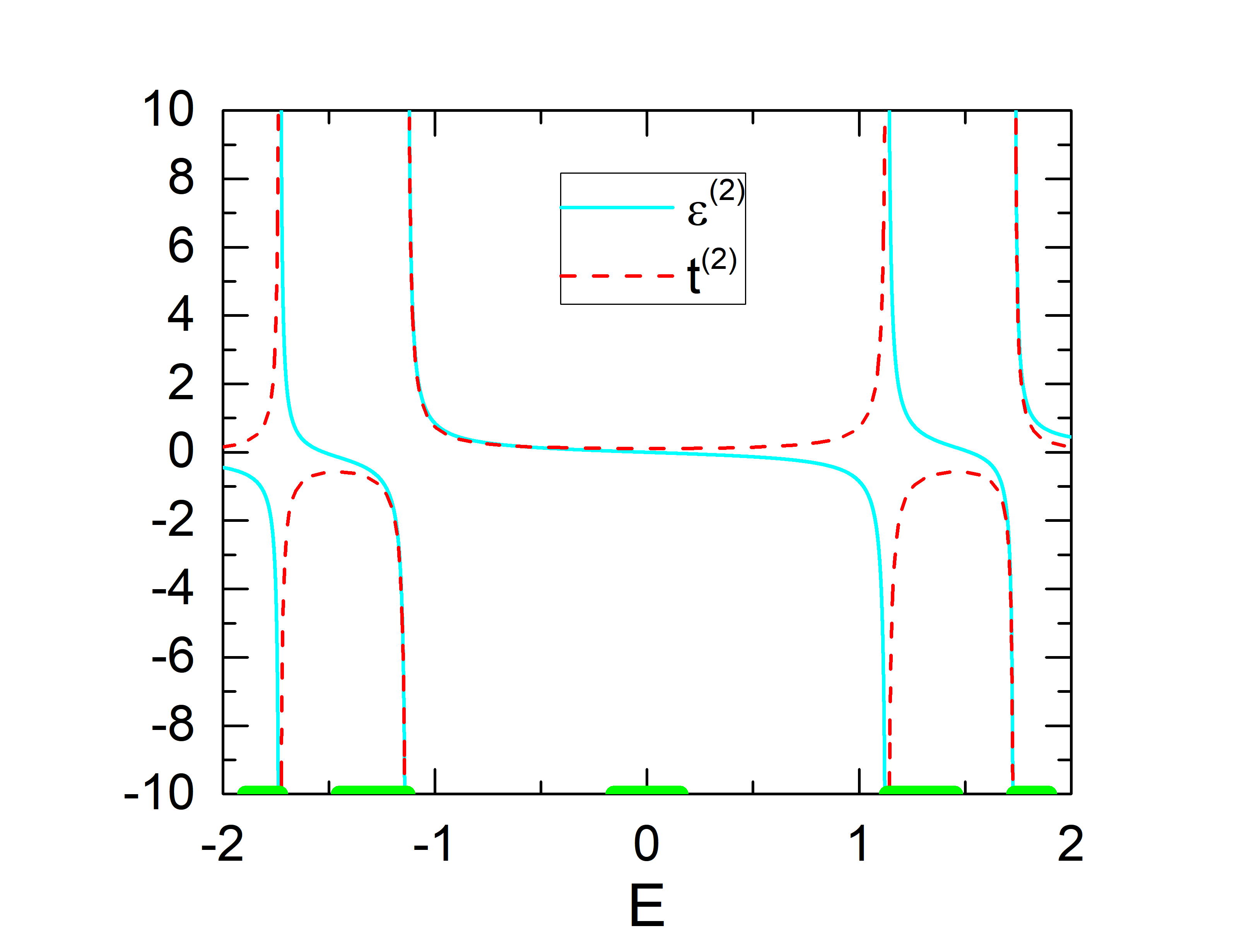}
            \label{fig_eff_Nw2} 
        \end{subfigure} 
    \end{minipage}
    \vspace{-15pt}
    \caption{ Effective renormalized description of domain-wall lattices. (a) Elementary three-site decimation. (b) Iterative reduction of a finite SSH segment containing $N_w$ hopping amplitudes $w$. (c) Effective periodic chain obtained after replacing each SSH segment by its renormalized counterpart. (d) Effective parameters for the elementary domain-wall lattice with $N_w=1$. (e) Effective parameters for the elementary domain-wall lattice with $N_w=2$. Red dashed curves correspond to the effective hopping $t^{(N_w)}$, while cyan solid curves denote the effective site energy $\varepsilon^{(N_w)}$. Green segments mark the energies satisfying the effective band condition, Eq.~(\ref{eq_band_condition}). } 
    \label{fig_renorm} 
\end{figure*}

Consider a three-site chain with site energies $\varepsilon_1$, $\varepsilon_2$, and $\varepsilon_3$, connected by hopping amplitudes $t_1$ and $t_2$, as shown in Fig.~\ref{fig_renorm}(a). The central site can be transformed into an equivalent two-site system characterized by the effective site energies and hopping parameters \cite{Farchioni1992}

\begin{subequations}
\label{eq:decimation}
\begin{align}
\varepsilon_L
&=
\varepsilon_1+\frac{t_1^2}{E-\varepsilon_2},
\\
\varepsilon_R
&=
\varepsilon_3+\frac{t_2^2}{E-\varepsilon_2},
\\
t_{\rm eff}
&=
\frac{t_1t_2}{E-\varepsilon_2}.
\end{align}
\end{subequations}

By iteratively applying this procedure to a finite SSH chain that begins and ends with a hopping amplitude $v$, the complete structure can be analytically renormalized to an effective two-site system, as illustrated in Fig.~\ref{fig_renorm}(b). For SSH segments containing $N_w$ hopping amplitudes $w$, the effective system is described by two identical site energies $\varepsilon^{(N_w)}$ connected through an effective hopping $t^{(N_w)}$.

The renormalized parameters can be obtained recursively through

\begin{equation}
A_1=E,
\qquad
B_1=E-\frac{w^2}{A_1},
\end{equation}

and

\begin{equation}
A_n
=
E-\frac{v^2}{B_{n-1}},
\qquad
B_n
=
E-\frac{w^2}{A_n}.
\end{equation}

The effective parameters are then

\begin{equation}
\varepsilon^{(N_w)}
=
\frac{v^2}{B_{N_w}},
\end{equation}

and

\begin{equation}
t^{(N_w)}
=
\frac{
v^{N_w+1}w^{N_w}
}{
A_1B_1A_2B_2\cdots A_{N_w}B_{N_w}
}.
\end{equation}

These expressions provide an exact description of the effective interaction between the domain-wall states localized at the ends of a finite SSH segment. Since the minibands discussed in Sec.~\ref{dws} are centered around $E=0$, it is particularly useful to examine the effective parameters in this limit. Taking $E\rightarrow0$, the effective site energy vanishes,

\begin{equation}
\varepsilon^{(N_w)}
\rightarrow 0,
\end{equation}
while the effective hopping reduces to

\begin{equation}
t^{(N_w)}
\rightarrow
v\left(\frac{v}{w}\right)^{N_w}.
\end{equation}

Because $v<w$ in the topological phase, the effective coupling decreases exponentially with increasing $N_w$. Consequently, domain-wall states separated by larger distances interact more weakly and generate progressively narrower minibands. This result explains the different miniband widths observed in Sec.~\ref{dws} for the $N_w=1$ and $N_w=2$ structures. It also provides a microscopic interpretation of the distinct effective couplings associated with the building blocks $W^{(1)}$ and $V^{(1)}$ introduced in Sec.~\ref{miniSSH}.

The effective description can also be used to interpret the band structures generated by periodic arrays of domain walls. Figure~\ref{fig_renorm}(c) shows the effective periodic chain obtained after replacing each SSH segment by its renormalized counterpart. Neighboring effective sites are connected through the hopping
\begin{equation}
    \label{eq:teff}
    t_{\text{eff}}=t^{(N_w)},
\end{equation}
while the onsite energy becomes
\begin{equation}
    \label{eq:epseff}
    \varepsilon_\text{eff}=2\varepsilon^{(N_w)},
\end{equation}
where the $2$ appears because every effective site is shared by two adjacent SSH segments.

If the effective parameters were independent of energy, the corresponding periodic chain would possess the dispersion relation

\begin{equation}
E
=
\varepsilon_\text{eff}
+
2t_\text{eff}\cos (ka) .
\end{equation}

Allowed bands would therefore exist whenever

\begin{equation}
\left|
E-\varepsilon_\text{eff}
\right|
<
2
\left|
t_\text{eff}
\right|.
\label{eq_band_condition}
\end{equation}

In the present case, however, both $\varepsilon_\text{eff}$ and $t_\text{eff}$ depend explicitly on the energy. Consequently, Eq.~(\ref{eq_band_condition}) must be interpreted self-consistently, and the allowed spectral regions are determined by the energy dependence of the effective parameters themselves.

Figures~\ref{fig_renorm}(d,e) show the effective parameters $\varepsilon^{(N_w)}$ and $t^{(N_w)}$ for the cases of $N_w=1$ and $2$, respectively. The green segments indicate the energies satisfying Eq.~(\ref{eq_band_condition}), using the effective parameters defined in \cref{eq:teff,eq:epseff}. Remarkably, these intervals coincide with the allowed bands observed in Fig.~\ref{fig3_PDW}, establishing a direct connection between the transport spectra, band structures and the effective renormalized description.

The structure of the effective bands is controlled by the singularities of the effective parameters. As demonstrated in Appendix~\ref{app_singularities}, each additional site absorbed during the decimation procedure generates exactly one new singularity, and the singularities of successive generations are interlaced. Furthermore, the effective site energies and hopping amplitudes share the same poles. The singularities partition the energy axis into a sequence of intervals. Within each interval, the effective parameters remain finite and continuous, allowing an independent branch of the effective band condition, Eq.~(\ref{eq_band_condition}), to be satisfied. Consequently, the singularities determine how the spectrum is partitioned into regions that can support allowed bands.

For $N_w=1$, two singularities partition the physical energy window into three intervals, each containing one allowed band. Similarly, the $N_w=2$ structure exhibits four singularities, producing five intervals and therefore five allowed bands. More generally, the number of allowed bands coincides with the number of intervals generated by the singularity structure.

\subsection{Renormalized description of hierarchical generations}

\begin{figure}[tb] 
    \centering 
    \includegraphics[width=\linewidth]{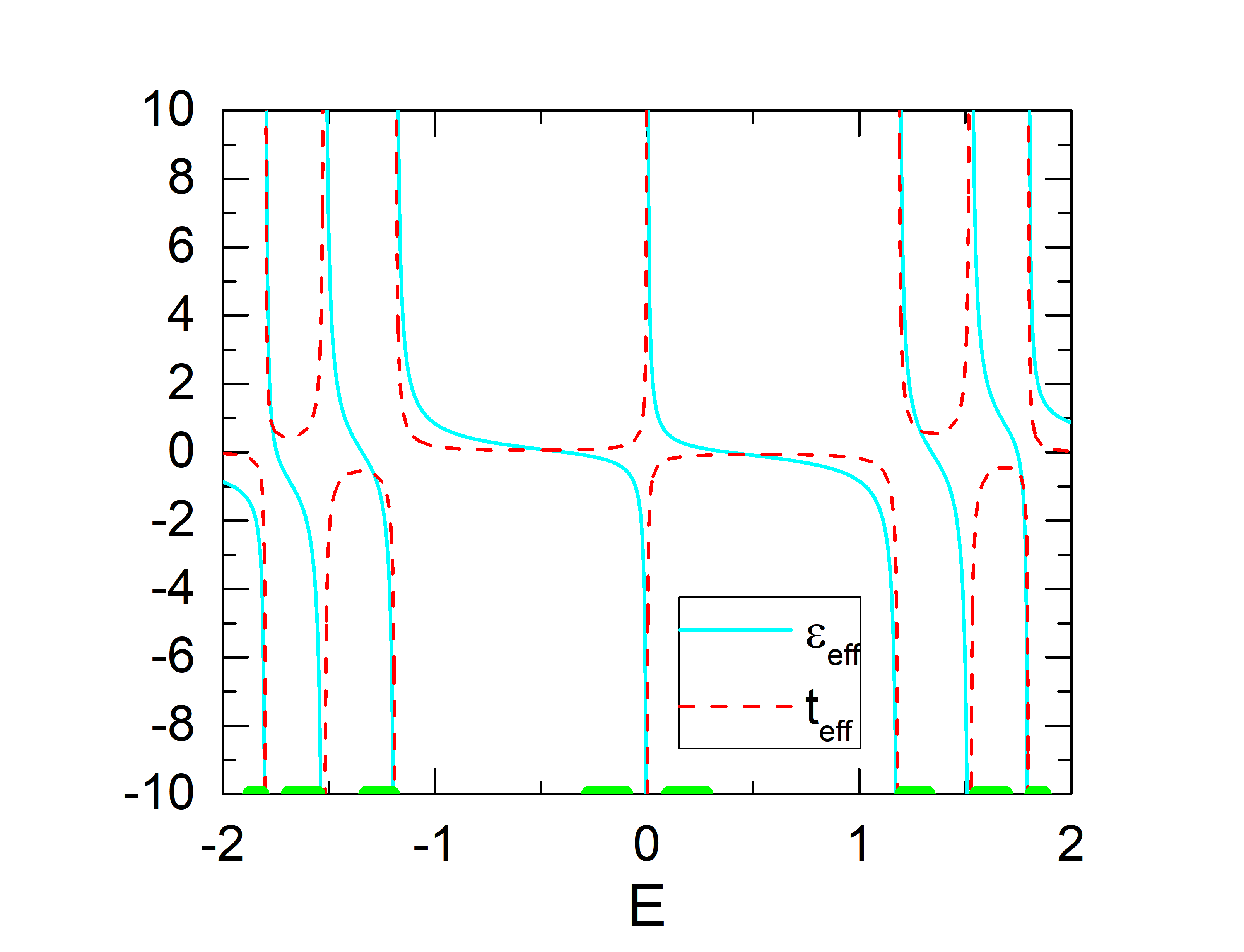} 
    \caption{ Effective parameters associated with the first-generation SSH miniband structure. Green lines indicate the energy intervals of transmission bands. The poles of the effective parameters delimit the allowed spectral regions.} 
    \label{fig_eff1} 
\end{figure}

The effective description can be extended naturally to the first-generation SSH miniband structures introduced in \cref{miniSSH}. In this case, the periodic unit cell is formed by concatenating one $V^{(1)}$ block and one $W^{(1)}$ block, corresponding to the sequence $vwvwv  vwv$. Following the decimation procedure of~\cref{eq:decimation}, we again obtain two effective sites connected by an effective coupling $t_\text{eff}$. Since the chain is not symmetric, the left and right effective site energies are generally different. Nevertheless, a periodic repetition of the unit cell produces a one-dimensional tight-binding chain of the type shown in Fig.~\ref{fig_renorm}(c), for which $\varepsilon_\text{eff}$ is given by the sum of the left and right effective energies. The resulting effective parameters of the periodic chain are shown in Fig.~\ref{fig_eff1}. Green segments indicate the energies satisfying the band condition of~\cref{eq_band_condition}.

\begin{figure}[tb] 
    \centering 
    \includegraphics[width=\linewidth]{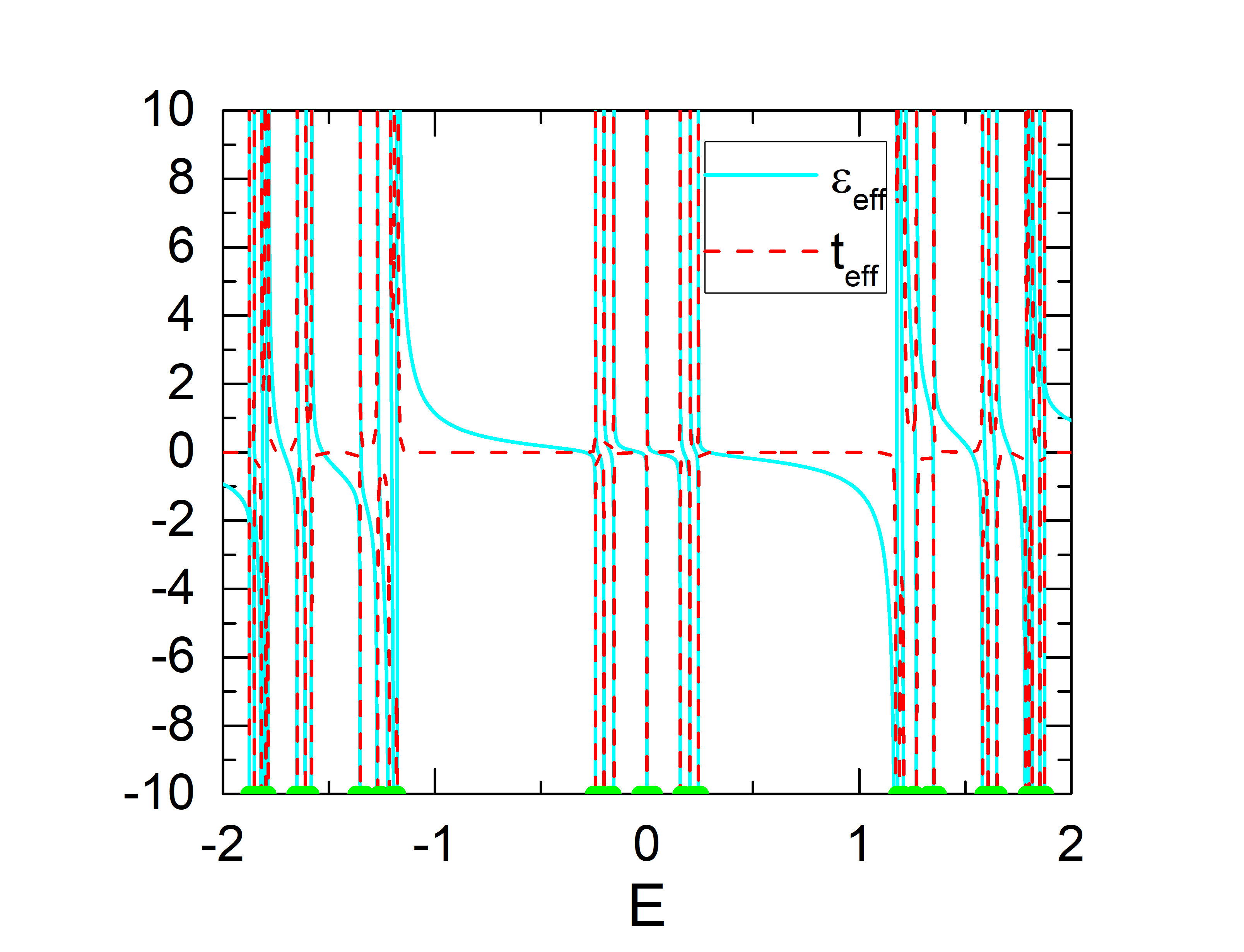} 
    \caption{ Effective parameters corresponding to the second hierarchical generation. Additional singularities appear between those of the previous generation, illustrating the interlacing property demonstrated in Appendix~\ref{app_singularities}. } 
    \label{fig_eff2} 
\end{figure}

Applying the same procedure once more for the concatenation of one $V^{(2)}$ block and one $W^{(2)}$ block produces the subsequent hierarchical generation of effective parameters, as shown in Fig. \ref{fig_eff2}.

In both cases, the energy intervals satisfying Eq.~(\ref{eq_band_condition}) coincide with the minibands observed in the corresponding transport spectra shown in the first and second columns of~\cref{fig4_res_Gens}. 

In all the examples analyzed in this work, each interval bounded by consecutive singularities supports exactly one allowed band. Most importantly, the hierarchy of singularities generated by the decimation procedure reproduces the same self-similar organization observed in the transport calculations. Additional singularities appear between those of the previous generation, refining the partition of the energy axis and creating new spectral regions in which the effective band condition can be satisfied. The recursive generation of singularities provides a natural interpretation of the progressive fragmentation of the spectrum and of the self-similar hierarchy displayed in Fig.~\ref{fig4_res_Gens}.

\section{Discussion} \label{disc} 

The results presented above reveal a progressive reorganization of the SSH spectrum across multiple hierarchical levels. At the first level, the introduction of domain walls generates localized defect states inside the SSH gap. As the number of domain walls increases, these states hybridize and form minibands whose spectral signatures are directly reflected in the transmission properties. The correspondence between the number of domain walls and the number of transmission resonances further demonstrates that the transport channels inside the original SSH gap originate from the collective interaction of the domain-wall states.

From this perspective, coupled domain-wall states may be regarded as emergent degrees of freedom defining an effective lattice embedded within the original SSH chain. Topological defects therefore cease to behave merely as localized excitations and instead become building blocks from which new spectral structures can be constructed. By controlling the arrangement of domain walls and the coupling between their associated states, it becomes possible to engineer effective transport channels and band structures within selected energy windows.

The SSH minibands discussed in Sec.~\ref{miniSSH} illustrate this idea explicitly. Two domain-wall lattices with different domain-wall separations generate distinct effective couplings between defect states. When these effective couplings are arranged according to an SSH pattern, the resulting miniband develops the same characteristic spectral organization observed in the original SSH model, including the formation of a gap, finite-size resonances, and the correspondence between transmission spectra and band structure. The emergence of an effective SSH structure within a miniband suggests that this construction can be iterated recursively.

The recursive structures presented in Sec.~\ref{recursive} show that this mechanism is not restricted to a single generation. As demonstrated by the effective renormalized description of Sec.~\ref{renorm}, each hierarchical level generates a new hierarchy of singularities in the effective parameters. These singularities progressively subdivide the energy axis into an increasing number of intervals, leading to the repeated formation of minibands and spectral gaps at smaller energy scales. In this way, new SSH-like spectra repeatedly emerge within progressively narrower energy intervals, producing a hierarchy of minibands that preserves the characteristic spectral fingerprint of the parent structure. In this picture, each hierarchical level generates new effective degrees of freedom that become the building blocks of the next one. The observed self-similarity therefore arises from the repeated hybridization of topological domain-wall states, the successive formation of effective SSH structures, and the recursive generation of interlaced singularities in the corresponding effective parameters.

Hierarchical and self-similar spectra are often associated with quasiperiodic systems and fractal geometries \cite{Dana2014,Song2026}. In the present case, however, the hierarchy emerges from a different mechanism. Rather than originating from quasiperiodic order in real space, the observed spectral hierarchy is generated through a recursive topological construction based on the interaction of domain-wall states. Topology thus acts not only as a mechanism for generating protected states but also as a mechanism for generating new effective degrees of freedom, transport channels, and spectral structures at progressively smaller energy scales.

Besides generating a hierarchy of spectral structures, the recursive construction also defines a substitution sequence of building blocks. Each recursive step replaces the symbols $W^{(n)}$ and $V^{(n)}$ with longer sequences of blocks from the previous generation, producing an inflation process analogous to those encountered in quasiperiodic systems.

For the particular recursive rule employed here, namely $N_W^{(k)}=1$ and $M_W^{(k)}=2$ at the intermediate levels, the asymptotic ratio between the lengths of the two building blocks approaches

\begin{equation}
    2+\sqrt{5}=\phi^{3}.
\end{equation}
This value corresponds to the cubic power of the golden ratio and characterizes the long-range organization generated by the recursive construction.

More generally, different choices of the recursive parameters generate a broader family of sequences associated with other metallic ratios, including the silver and copper ratios \cite{sanchez_renormalization_2016}. This observation suggests a natural connection between the hierarchical topological construction presented here and the spectral properties commonly encountered in quasiperiodic systems.

The effective renormalized description developed in Sec.~\ref{renorm} further suggests that the recursive spectral hierarchy can be interpreted as a hierarchy of effective parameters and their associated singularities. Within this picture, additional spectral gaps emerge whenever new singularities are generated, while the intervals bounded by consecutive singularities support the corresponding minibands.

The construction introduced here provides a systematic route for generating multiscale spectral structures and hierarchies of effective parameters from topological domain-wall states. Beyond the specific examples considered in this work, the recursive framework could be extended to other topological lattices, higher-dimensional systems, or engineered platforms where localized topological states can be selectively coupled and organized into hierarchical patterns across multiple length and energy scales.

\section{Conclusions} \label{conclusions}

We have investigated the transport properties of SSH chains containing multiple domain walls and demonstrated that the interaction of their associated localized states gives rise to a hierarchy of emergent spectral structures. The transport calculations reveal that each domain wall contributes a localized state inside the SSH gap and that the hybridization of these states produces minibands whose signatures are directly reflected in the transmission spectra.

By combining domain-wall lattices with different domain-wall separations, we constructed effective SSH structures within the miniband subspace. The resulting transmission spectra exhibit the same characteristic features found in conventional SSH chains, including the formation of transmission gaps, finite-size resonances, and the correspondence between transport spectra and band structures. In this sense, the minibands generated by domain-wall states develop an emergent SSH structure at a smaller energy scale.

The recursive application of this construction produces successive generations of effective SSH structures. As a consequence, effective SSH transmission spectra repeatedly reappear within progressively narrower energy intervals. The effective renormalized description reveals that this process is accompanied by the recursive generation of interlaced singularities in the effective parameters, which progressively subdivide the energy axis and produce new minibands and spectral gaps. The resulting spectra display a clear self-similar hierarchy in which the characteristic gap-and-resonance structure of the SSH model is preserved across multiple scales.

These results show that topological domain-wall states can be regarded as emergent degrees of freedom from which new transport channels, effective couplings, and band structures can be engineered. More generally, the hierarchical framework introduced here provides a systematic route for constructing multiscale spectral structures through the controlled hybridization of topological defect states. The associated hierarchy of effective parameters and singularities offers a complementary perspective on the emergence of self-similar spectra in topological systems, opening new possibilities for the design of transport properties in topological lattices.

\section*{Acknowledgments} 
This work was supported by UNAM-PAPIIT Grant No. IN116025. Computations were performed at Miztli under Project LANCAD-UNAM-DGTIC-329. C.G.G.P. wishes to acknowledge the resources from project LANCAD 29-2026 of the National Autonomous University of Mexico. A. P. thanks the support of a Postdoctoral grant from Secretaría de Ciencia, Humanidades, Tecnología e Innovación (SECIHTI).

\appendix
\renewcommand{\theequation}{A\arabic{equation}}
\setcounter{equation}{0}

\section{Generation of singularities under successive decimation} \label{app_singularities} 

In this appendix, we analyze the singularities generated by the successive decimation of a one-dimensional tight-binding chain.

Consider a chain composed of $N$ sites where site $i$ has on-site energy $\varepsilon_i$ and neighboring sites are connected through hopping amplitudes $t_i$. Starting from the left edge, the decimation procedure successively eliminates intermediate sites and generates energy-dependent effective parameters. To study the proliferation of singularities, we first consider the effective on-site energy of the rightmost site. The remaining effective parameters are discussed afterwards.

\subsection{Right effective energies}

The first decimation step yields

\begin{equation}
    \varepsilon_R^{(1)} = \varepsilon_3+ \frac{t_2^2}{E-\varepsilon_2}.
\end{equation} 
This function possesses a single singularity at 
\begin{equation} 
    E=\varepsilon_2. 
\end{equation}
Successive decimation steps generate a sequence of effective site energies. After the $n$-th decimation step, the effective energy satisfies
\begin{equation}
    \varepsilon_R^{(n)} = \varepsilon_{n+2} + \frac{t_{n+1}^2} {E-\varepsilon_R^{(n-1)}}. 
    \label{eq_recursive_eps}
\end{equation}
The singularities of $\varepsilon_R^{(n)}$ occur whenever the denominator of Eq.~(\ref{eq_recursive_eps}) vanishes, namely when

\begin{equation}
    E=\varepsilon_R^{(n-1)}(E).
    \label{eq_fixedpoint}
\end{equation}
Thus, the generation of singularities is completely determined by the fixed points of the effective energy from the previous decimation step. The problem therefore reduces to counting the solutions of Eq.~(\ref{eq_fixedpoint}).

We first establish the monotonic character of the effective energies. Differentiating Eq.~(\ref{eq_recursive_eps}) gives

\begin{equation}
    \frac{d\varepsilon_R^{(n)}}{dE} = -\frac{t_{n+1}^2} {\left(E-\varepsilon_R^{(n-1)}\right)^2} \left( 1-\frac{d\varepsilon_R^{(n-1)}}{dE} \right).
\end{equation}
For $n=1$,

\begin{equation}
    \frac{d\varepsilon_R^{(1)}}{dE} = -\frac{t_2^2} {(E-\varepsilon_2)^2} <0. 
\end{equation}
Assuming that

\begin{equation}
    \frac{d\varepsilon_R^{(n-1)}}{dE}<0 
\end{equation}
within every interval where the function remains finite, we obtain

\begin{equation}
    1-\frac{d\varepsilon_R^{(n-1)}}{dE}>0, 
\end{equation}
which immediately implies
    
\begin{equation}
    \frac{d\varepsilon_R^{(n)}}{dE}<0. 
\end{equation}

By induction, $\varepsilon_R^{(n)}$ is strictly decreasing within each interval between consecutive singularities.

We now determine the number of singularities generated at each iteration. Suppose that $\varepsilon_R^{(n-1)}$ possesses $n-1$ singularities located at

\begin{equation}
    s^{(n-1)}_1 < s^{(n-1)}_2 < \cdots < s^{(n-1)}_{n-1}.
\end{equation}

These singularities divide the real axis into $n$ intervals,

\begin{equation} 
    I_1 = (-\infty,s^{(n-1)}_1), 
\end{equation} 
\begin{equation}
    I_k = \left( s^{(n-1)}_{k-1}, s^{(n-1)}_{k} \right), \qquad k=2,\ldots,n-1, \end{equation}
and
\begin{equation}
    I_n = (s^{(n-1)}_{n-1},\infty).
\end{equation}

Within each interval $I_k$, $\varepsilon_R^{(n-1)}$ is continuous and strictly decreasing. Furthermore, the effective energy diverges with opposite signs on both sides of every singularity,

\begin{equation}
    \lim_{E\to (s_i^{(n-1)})^-} \varepsilon_R^{(n-1)}(E) = -\infty
\end{equation}
while
\begin{equation}
    \lim_{E\to (s_i^{(n-1)})^+} \varepsilon_R^{(n-1)}(E) = +\infty. 
\end{equation}

Defining
\begin{equation}
    f(E)=\varepsilon_R^{(n-1)}(E)-E, 
\end{equation}
it follows from the previous limits that $f(E)$ changes sign across every interval. Since $f(E)$ is continuous between consecutive poles, the intermediate value theorem guarantees the existence of at least one solution of

\begin{equation}
    f(E)=0.
\end{equation}

Moreover,
\begin{equation}
    f'(E) = \frac{d\varepsilon_R^{(n-1)}}{dE} - 1 < 0,
\end{equation}
which implies that $f(E)$ is strictly decreasing. Therefore, each interval contains exactly one solution of Eq.~(\ref{eq_fixedpoint}). 

Since there are $n$ intervals, Eq.~(\ref{eq_fixedpoint}) possesses exactly $n$ solutions. Hence, the effective energy $\varepsilon_R^{(n)}$ contains exactly $n$ singularities, which we denote by

\begin{equation}
    s^{(n)}_1 < s^{(n)}_2 < \cdots < s^{(n)}_n.
\end{equation}

Because each interval contains exactly one solution of Eq.~(\ref{eq_fixedpoint}), every singularity of generation $n$ is located within an interval bounded by singularities of generation $n-1$. For each singularity $s_i^{(n-1)}$ of generation $n-1$, 

\begin{equation} 
    s^{(n)}_i < s^{(n-1)}_i < s^{(n)}_{i+1}, \qquad i=1,\ldots,n-1. 
\end{equation} 
Hence, the singularities of successive generations are interlaced.
    
We therefore conclude that every additional hopping absorbed during the decimation procedure generates exactly one additional singularity in the right effective energy. Moreover, the singularities produced at successive iterations form an interlaced hierarchy along the energy axis. Each decimation step increases the number of intervals separating consecutive singularities by one, progressively refining the partition of the energy axis.

\subsection{Effective hopping singularities} We now show that the effective hopping amplitudes generated during the decimation procedure possess exactly the same singularities as the effective energies studied above.

Let $t_{\rm eff}^{(n)}$ denote the effective hopping obtained after the $n$-th decimation step. From the decimation rule,

\begin{equation}
    t_{\rm eff}^{(n)} = \frac{ t_{n+1}\, t_{\rm eff}^{(n-1)} } { E-\varepsilon_R^{(n-1)} }, 
    \label{eq_teff_recursive}
\end{equation} 
where $\varepsilon_R^{(n-1)}$ is the effective energy introduced in Eq.~(\ref{eq_recursive_eps}). 

At first sight, Eq.~(\ref{eq_teff_recursive}) suggests that the singularities of $t_{\rm eff}^{(n)}$ arise both from the poles already present in $t_{\rm eff}^{(n-1)}$ and from the singular behavior of the factor $E-\varepsilon_R^{(n-1)}$. However, these two contributions are not independent. 

From the previous section, the singularities of $\varepsilon_R^{(n-1)}$ satisfy 
\begin{equation}
    E=\varepsilon_R^{(n-2)}(E).
\end{equation} 
These are precisely the singularities inherited by $t_{\rm eff}^{(n-1)}$. Therefore, if $s_i^{(n-1)}$ denotes a singularity of generation $n-1$, then in the vicinity of $s_i^{(n-1)}$,
\begin{equation} 
    t_{\rm eff}^{(n-1)} \sim \frac{A_i}{E-s_i^{(n-1)}}, 
\end{equation} 
while 
\begin{equation} 
    E-\varepsilon_R^{(n-1)}(E) \sim \frac{B_i}{E-s_i^{(n-1)}}. 
\end{equation} 

Therefore, 
\begin{equation} 
    t_{\rm eff}^{(n)} \sim \frac{A_i}{E-s_i^{(n-1)}} \, \frac{E-s_i^{(n-1)}}{B_i} = \frac{A_i}{B_i}, 
\end{equation} 
which is finite.

Thus, the poles inherited from $t_{\rm eff}^{(n-1)}$ are exactly cancelled by the singular denominator introduced at the next decimation step.

The only singularities that survive in $t_{\rm eff}^{(n)}$ are therefore the solutions of

\begin{equation}
    E=\varepsilon_R^{(n-1)}(E), 
\end{equation}
which are precisely the singularities of $\varepsilon_R^{(n)}$. 

Consequently, the effective hopping and effective site energies possess the same set of singularities at every decimation step, 

\begin{equation} 
    \mathrm{Poles}\!\left(t_{\rm eff}^{(n)}\right) = \mathrm{Poles}\!\left(\varepsilon_R^{(n)}\right). 
\end{equation}
The interlacing property demonstrated previously for $\varepsilon_R^{(n)}$ therefore applies equally to the effective hopping amplitudes. As a consequence, all effective parameters generated by the decimation procedure share the same hierarchical structure of singularities.

\subsection{Left effective energies}
The previous results can be extended to the effective energy of the leftmost site without requiring an explicit analysis of its recursion relation. The decimation procedure may be performed equivalently from right to left. In that case, the role played previously by $\varepsilon_R^{(n)}$ is assumed by the effective energy of the leftmost site, denoted here by $\varepsilon_L^{(n)}$. Since the derivation presented above depends only on the recursive structure of the decimation procedure and not on the direction in which the sites are eliminated, the same proof applies directly to $\varepsilon_L^{(n)}$.

Furthermore, the final two-site effective Hamiltonian obtained after decimating a given chain must be independent of whether the intermediate sites are eliminated from left to right or from right to left. In particular, the resulting effective hopping amplitude $t_{\rm eff}^{(n)}$ is unique and therefore possesses the same singularities in both constructions. Because the singularities of $t_{\rm eff}^{(n)}$ coincide with those of $\varepsilon_R^{(n)}$, and the right-to-left decimation produces the same effective hopping but associates its singularities with $\varepsilon_L^{(n)}$, it follows that \begin{equation} \mathrm{Poles}\!\left(\varepsilon_L^{(n)}\right) = \mathrm{Poles}\!\left(\varepsilon_R^{(n)}\right) = \mathrm{Poles}\!\left(t_{\rm eff}^{(n)}\right). \end{equation} 

Hence, all effective parameters generated by the decimation procedure share the same hierarchy of singularities. Consequently, the recursive generation of interlaced singularities is a universal feature of the decimation procedure and does not depend on the particular effective parameter under consideration.

\bibliographystyle{unsrt}
\bibliography{Bibliography.bib}
\end{document}